\def\year{2021}\relax
\documentclass[letterpaper]{article} 
\usepackage{aaai21}  
\usepackage{times}  
\usepackage{helvet} 
\usepackage{courier}  
\usepackage[hyphens]{url}  
\usepackage{graphicx} 
\usepackage{natbib}  
\usepackage{caption} 
\usepackage{colortbl}
\usepackage{tabulary}
\usepackage{etoolbox}

\usepackage{subcaption}
\usepackage{enumitem}
\usepackage{diagbox}
\usepackage{booktabs}
\usepackage{amsmath}

\usepackage{color}

\usepackage{pifont}

\newcommand{\gp}{\textit{TGP }}

\begin{document}


\author{Zhouhan Chen, Juliana Freire, Jonathan Nagler, Joshua A. Tucker \\ New York University \\ Haohan Chen, The University of Hong Kong \\ {zhouhan.chen@nyu.edu}
}


\title{Understanding how people consume low quality and extreme news using web traffic data}
\maketitle

\begin{abstract}
To mitigate the spread of fake news, researchers need to understand who visit fake new sites, what brings people to those sites, where visitors come from, and what content they prefer to consume. In this paper, we analyze web traffic data from \textit{The Gateway Pundit} (\textit{TGP}), a popular far-right website that is known for repeatedly sharing false information that has made its web traffic available to the general public. We collect data on 68 million web traffic visits to the site over a month period and analyze how people consume news via multiple features. Our traffic analysis shows that search engines and social media platforms are main drivers of traffic; our geo-location analysis reveals that~\gp is more popular in counties that voted for Trump in 2020; and our topic analysis shows that conspiratorial articles receive more visits than factual articles. 

Due to the inability to observe direct website traffic, existing research uses alternative data source such as engagement signals from social media posts. To validate if social media engagement signals correlate with actual web visit counts, we collect all Facebook and Twitter posts with URLs from~\gp during the same time period. We show that all engagement signals positively correlate with web visit counts, but with varying correlation strengths. Metrics based on Facebook posts correlate better than metrics based on Twitter. Our unique web traffic data set and insights can help researchers to better measure the impact of far-right and fake news URLs on social media platforms.

\end{abstract}


\section{Introduction}
\label{section:introduction}
Fake news site is a major threat on today's Internet~\cite{persily_tucker_2020, fakenews-2016-election, spread-of-true-and-false-news}. How to measure the consumption of fake news URLs remains a challenge. Since there is no single metric to quantify the spread of information, the choice of metrics can affect downstream analysis and alter final conclusions. There are two approaches to measuring fake news consumption: indirect and direct. 

For indirect measurement, a common method is to collect social media posts containing the URL of interest, calculate engagement signals, and use those metrics as a proxy for URL popularity~\cite{fakenews-2020-election, less-than-you-think, cracking-open-the-news-feed}. Indirect measurements reveal how people \textit{share} news URLs, but not how people actually \textit{visit} those URLs~\cite{fake-news-not-antitrust}. 

Only a few studies use direct measurement data. For example,~\cite{rise-and-fall-of-fake-news-sites} collects visit data to fake news sites from third party services such as SimilarWeb and CheckPageRank to assess user engagement. In another study, ~\cite{microsoft-measurement} gathers browsing data from Microsoft Internet Explorer and Edge to analyze visiting patterns to fake news sites before the 2016 US Election. As far as we know, one unexplored data source is web traffic data collected on the server side. This type of web traffic data has rich features that alternative sources do not have. Even though most news websites record their traffic, few make the data publicly available. 


During an audit of popular far-right and extreme news websites, we discovered that~\gp makes its website traffic available to the general public.~\gp is one of the top three right-wing news sites with the largest percentage of traffic surge from December 2019 to December 2020~\cite{surge-in-traffic}. It is also one of ``the top-three most cited domains in tweets spreading false and misleading narratives about voter fraud in 2020''~\cite{fakenews-2020-election}. Even though -- or perhaps because -- the site constantly shares misinformation~\cite{The-Gateway-Pundit-NewsGuard-Nutrition-Label, partisanship-propaganda-disinformation}, it remains highly influential. For example, its articles were cited by Former President Trump’s lawyer and referenced in Trump’s Impeachment Defense Memo~\cite{trail-memorandum}. All of these features make~\gp an ideal case study to understand online extreme news consumption behavior.


Given this opportunity, we crawl the entire web traffic from~\gp for one month from February 4, 2021 to March 3, 2021. We collect a total of 68 million website visits. Our analysis is two-fold: we first explore available features within the web traffic data to understand how people consume low quality news; we then collect additional social media posts to test correlations between social media engagement signals and actual web visit counts. Our substantive findings include:

\begin{enumerate}[leftmargin=*]
    \item Search engines such as Google, Duckduckgo and Bing account for 88.5\% of external referral traffic to~\gp home page. Social media platforms including Twitter, Facebook, Telegram and Gab account for more than 42\% of external referral traffic to~\gp article pages.
    \item At the county level,~\gp is more popular in counties that vote for Trump. At the state level,~\gp is more popular in ``swing states'' such as Georgia and Arizona.
    \item Topic modeling reveals that articles that mention ``2020 US election fraud'' are visited by 29\% more users, and spread 8 hours longer, compared with articles of other topics. Those viral articles usually cover events happening in swing states. 
    \item Social media engagement signals positively correlate with actual website visit counts. Not all metrics are the same: Facebook metrics achieve a stronger correlation than Twitter metrics.
\end{enumerate}

To the best of our knowledge, our work is the first to analyze server-side web traffic data of a popular low quality news site, and the first to correlate social media engagement signals with actual web traffic counts. In the future, we plan to apply our method to similar server-side web traffic data, although getting access to additional data sets remains challenging. 




\section{Method}
\label{section:method}
In this section, we first explain how we collect the entire visit traffic from~\gp for one month. We then give an overview of the collected data, and address issues related to data integrity, missing data and data privacy.

\subsection{Data collection}
\gp uses StatCounter, a web traffic service, to capture visitor traffic. To access traffic data to~\gp, users can either visit a publicly available web portal, or download the data by sending an HTTP GET request to a URL endpoint, which we refer to as the ~\textit{download URL}. Users need to specify two parameters in the URL, which we refer to as \textit{StartTime} and \textit{EndTime}.\footnote{The web portal is available at ~\url{https://statcounter.com/p9449268/summary/?guest=1}. The~\textit{download URL} follows the following pattern: \textit{\url{https://statcounter.com/p9449268/csv/download_log_file?range=StartTime--EndTime}}. \textit{StartTime} and \textit{EndTime} must be in ISO format, such as \textit{2021-04-12T02:18:41}. For a formal definition of ISO format, refer to \url{https://www.w3.org/TR/NOTE-datetime}.}

During our testing phase, we find that no matter what \textit{StartTime} and \textit{EndTime} we set, the downloaded CSV file always contains traffic captured during the most recent 20 minutes. To collect website traffic continuously, we set up a Selenium Chrome Browser to visit the~\textit{download URL} every 15 minutes, from February 3, 2021 to March 3, 2021. We choose a 15 minute interval because it is below the 20-minute interval with a safe margin. One side effect is that our data has duplicates. To remove duplicates, we identify that each website visit is uniquely defined by the combination of five features: \textit{datetime, url, ip, os}, and \textit{browser}. Therefore, we only keep the first record if multiple records have the same five-feature combination.


\begin{table}
\begin{tabular}{p{0.17\linewidth}p{0.4\linewidth}p{0.45\linewidth}}
\toprule
   feature & example & description \\
\midrule
 datetime & 2021-02-02 18:23:31	 & date and time of the visit \\
 ip & 10.11.123.12 &      IPv4 adddress \\
 os & IOS &     operating system \\
 url & thegatewaypundit.com	 & url visited \\
 isp & Verizon & Internet service provider \\
 country & USA & country of IP \\
 city & Houston & city of IP \\
 region & Texas & region \\
 referrer & google.com & previous url \\
 page title & Expert claims... & title of the article \\
 browser & Safari & browser \\
 resolution & 375 $\times$ 667 & resolution \\
\bottomrule
\end{tabular}
\caption{\label{table:overview-datasett}Features in~\gp web traffic data set.}
\end{table}

\subsection{Data integrity}
\begin{figure}[!h]
  \centering
  \includegraphics[width=\linewidth]{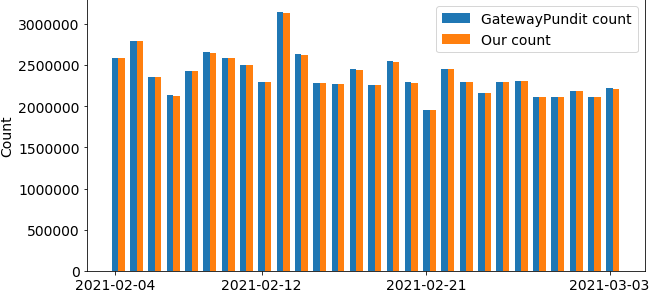}
  \caption{Total number of visits per day. Blue bar is the official count from~\textit{TGP}, and red bar is the count from our data set. Our data collection has a completeness ratio of more than 99.8\% on a daily basis.}
  \label{fig:data-integrity}
\end{figure}

To validate that our collection method captures the entire traffic, we compare the daily number of visits reported by Statcounter against the number calculated from our collection after de-duplication.~\footnote{The official aggregated counts is accessible from the following Statcounter URL:~\url{https://statcounter.com/p9449268/summary/daily-pur-labels-bar-20210204_20210303/}} Figure~\ref{fig:data-integrity} shows that our data set has a completeness ratio of more than 99.8\% on a daily basis. We define the completeness ratio as our number of visits divided by Statcounter's number of visits. The lost entries are possibly caused by parsing errors or corrupted network packages. We believe that this small number of missing entries (less than 0.2\%) will not affect trends we observe.

\subsection{Missing data and bot traffic}
Even though we capture the entire web traffic, our data source (StatCounter) has several inherent problems. One potential issue is under-counting. For example, anyone who blocks HTTP and HTTPS request to StatCounter will not have their visits logged by the server. This can happen if people install certain anti-tracking plug-ins. Unfortunately, it is impossible to know exactly how many users install anti-tracking tools, as those tools are designed to hide web visit history.

Another problem is the presence of bot traffic. Bots are programs that automatically visit web pages. According to the documentation, StatCounter does not record most bots or crawlers, because clients have to actually load javascript for their hit to be logged in the system~\cite{statcounter-ignore-crawler}. For more advanced bots that emulate human behavior (load javascript, click buttons), there is no way to distinguish their traffic from real human traffic. To sum up, even though the amount of missing data and advanced bot traffic is unknown and undetectable, we believe that those irregularities will not affect the overall trend during our analysis.

\subsection{Data privacy}
To address concerns regarding data privacy, we first note that our web traffic data does not contain any personal identifiable information such as name, phone number, cookie, session ID, device ID, email address, etc,. Additionally, all of our results presented below are aggregated. 

\subsection{Robustness of findings}
Despite the challenges noted above associated with our data collection process, two factors give us confidence in the robustness of our findings. First, our data is about as close to the ground truth as one can hope for in any sort of online data analysis, with a 99.8\% completeness rate. Second, we collect more than 68 million page visits that span a full month. This extended period of time ensures that any daily or hourly data irregularity is smoothed out and the overall trend preserved.





\section{Insights from 68 million web visits}
\label{section:result}
In this section, we take a multi-pronged approach to analyze our one-month web visit data along multiple dimensions. To better understand how people consume low quality news, we start by visualizing when people visit the site and from what type of device. We then analyze referrer links to understand which sites bring users to~\gp. We also leverage geo-spatial information to validate if people who visit~\gp come from areas that voted more favorably for Donald Trump in the 2020 Presidential Election. Finally, we apply topic clustering techniques to quantify what topics are discussed, and what topics are more likely to go viral.


\subsection*{Finding 1: The majority of users visit the site during the day on mobile devices.}
\begin{figure}[!h]
  \centering
  \includegraphics[width=\linewidth]{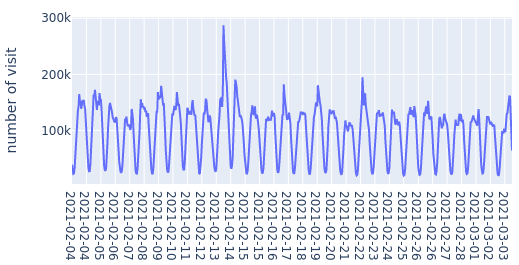}
  \caption{Number of visits per hour, from February 4, 2021 to March 3, 2021. There is a peak on February 13, 2021, the day Donald Trump was acquitted on impeachment charges. Our data shows that the two most visited articles during that day both covered this event.}
  \label{fig:visit-per-hour}
\end{figure}

Our data collection contains 68,268,818 unique visits, from February 3, 2021 to March 3, 2021. Figure~\ref{fig:visit-per-hour} plots the number of visits per hour. Since more than 95\% of the visits come from the United States, we see a regular and circadian pattern where the traffic increases during the day, and decreases during the night. The daily peak hourly visit is around 200,000. The only exception is one hour in February 13, 2021, with a recorded visit of nearly 300,000. February 13, 2021 is the day Donald Trump was acquitted on impeachment charges. After checking the data set, we find that the two most visited articles published that day are both about impeachment charges.\footnote{The two articles are \url{https://www.thegatewaypundit.com/2021/02/breaking-senate-votes-57-43-acquit-donald-trump-seven-republicans-voted-convict/} and \url{https://www.thegatewaypundit.com/2021/02/breaking-trump-releases-statement-following-impeachment-acquittal/}}

\begin{figure}[!h]
  \centering
  \includegraphics[width=\linewidth]{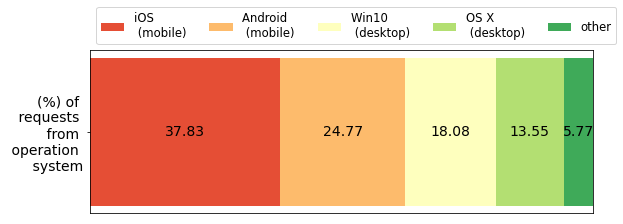}
  \caption{Most used operating systems among~\gp visitors. 80\% of visits come from mobile devices.}
  \label{fig:top-10-os}
\end{figure}

To understand how people visit~\textit{TGP}, we look at the operating systems (OS) column as it reveals what device people use. According to Figure~\ref{fig:top-10-os}, more than 80\% of visits come from mobile operating systems including IPhone and Android devices. If this finding holds for other low quality news sources, it suggests that research that mostly focuses on desktop users may miss a large proportion of the population that visits low quality news sources~\cite{misinformation-in-action}.

\subsection*{Finding 2: Search engines and social media sites are the main drivers of traffic to \textit{TGP}}
Knowing what websites bring people to~\gp helps us to identify the source of traffic and to design intervention strategies to slow down the spread of fake news. To reconstruct traffic flows, we use the \textit{referrer} column in our web traffic data. When a browser navigates to URL $B$ from URL $A$, it usually includes a string called referrer in the HTTP request (in our example $A$ is the referrer of $B$). Among 68,268,818 visits, 35,296,042 (52\%) have referrers. For visits that do not have referrers, either users visit a URL directly, or the browser strips the referrer, which can happen when certain privacy-enhancing features are turned on~\cite{mozilla-referrer}. To aggregate referrers that belong to the same site, we normalize each referrer URL to its domain name, removing hostname, path, and other query parameters. We consider two referral behaviors based on the destination URL: sites that bring users to the home page, and sites that bring users to an article page. A \textbf{home page} URL points to domain \textit{thegatewaypundit.com}, while an \textbf{article page} URL has the form \textit{thegatewaypundit.com\textbackslash ARTICLE}. Each type of traffic flow has its own characteristics, which we analyze separately.\\

\begin{figure}[!h]
  \centering
  \includegraphics[width=\linewidth]{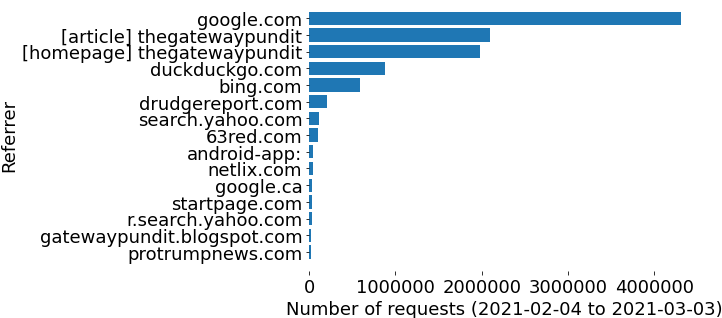}
  \caption{Top 15 domains that bring users to the home page.}
  \label{fig:referrer-homapage}
\end{figure}

\noindent
\textbf{Websites that bring users to the home page.}
Figure~\ref{fig:referrer-homapage} shows the top 15 domains that bring visitors to \gp home page. Three major search engines (Google, Duckduckgo and Bing) account for 88.5\% of external referral traffic. Among them,~\textit{Google.com} is the top driver of home page traffic (66\%). The anonymous search engine~\textit{duckduckgo.com} is the fourth (13\%), and the Microsoft-developed \textit{bing.com} ranked the fifth (9\%).\footnote{This finding is likely indicative of the role that search engines play in people's quest for political information generally, although it would be interesting to see if these percentages were similar for a mainstream news publication such as \textit{The New York Times}.} 

Second to Google are internal~\textit{TGP} article pages. This shows when people browse articles on~\textit{TGP}, they usually navigate back to the home page from different article pages. The third referrer is~\gp home page. This is likely caused by people clicking links to the home page when they are already at the home page. Further down the list are far-right and conservative news sites such as \textit{drudgereport.com}, \textit{63red.com} and \textit{protrumpnews.com}. 

We also identify referrers from suspected phishing domains. One such domain is \textit{\url{netlix.com}}, ranked number ten. The domain name used to be a center of a lawsuit. According to a legal complaint filed by Netflix in 2009, the video streaming company claims that the domain name ``netlix'' looks too similar to ``netflix,'' and requests \textit{\url{netlix.com}} to be transferred to Netflix.~\footnote{\url{https://www.adrforum.com/domaindecisions/1287043.htm}} The court rejected the order, and~\textit{\url{netlix.com}} still belongs to its original owner. As of September 14, 2021, the website does not host actual content, but automatically redirects users to~\textit{TGP}. We do not know what motivates the owner of \textit{\url{netlix.com}} to redirect visitors to~\textit{TGP}. We will keep monitoring the site as previous research demonstrates that URL redirection is a common technique to distribute unwanted or malicious software~\cite{url-redirection-campaign}.\\

\begin{figure}[!h]
  \centering
  \includegraphics[width=\linewidth]{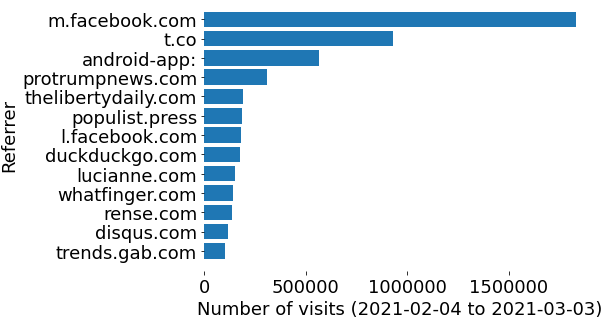}
  \caption{Top 15 domains that bring users to an article page. The top two referrers are~\gp home page and~\gp article pages. Together they account for more than 80\% of referral traffic. We drop those two domains for better visualization.}
  \label{fig:referrer-article}
\end{figure}

\noindent
\textbf{Websites that bring users to an article page.}
Figure~\ref{fig:referrer-article} shows the top 15 domains that bring visitors to an article page. The top two referrers -- home page and article page -- are both internal traffic. This indicates that (a) most users first land on the home page before clicking an individual article, (b) some users click a new article page while browsing an existing article page, since different articles are interlinked together. When we exclude referrers from~\url{thegatewaypundit.com}, we can classify the rest of sites into two groups:

\begin{enumerate}[leftmargin=*]
    \item \textbf{Social media platforms} including Twitter, Facebook, and emerging platforms such as Telegram and Gab. Together they account for 42\% of external referral traffic.
    \item \textbf{Conservative news sites} such as \textit{protrumpnews.com}, \textit{thelibertydaily.com}, \textit{populist.press} and \textit{whatfinger.com}. Those sites repost articles from~\gp on a regular basis.
\end{enumerate}

\begin{figure}[!h]
  \centering
  \includegraphics[width=\linewidth]{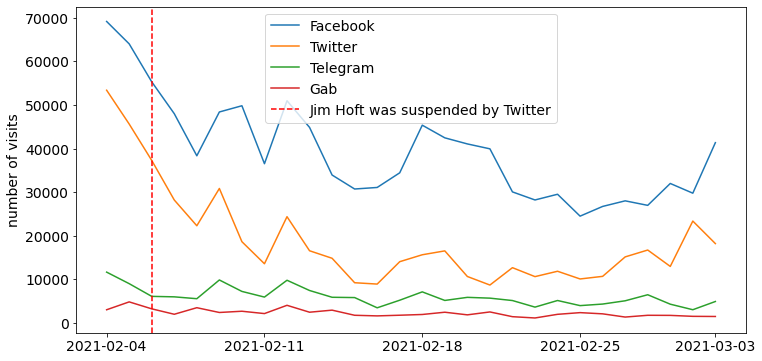}
  \caption{Daily referrers by platform. Facebook and Twitter are the top two traffic drivers, although their referral traffic volumes started to decline since February 6, 2021, when Jim Hoft, founder of~\gp, was suspended by Twitter.}
  \label{fig:daily-referrer-platform}
\end{figure}

To further understand how much role each social media platform plays in driving the traffic, we plot the daily number of visits with referrers from four different social media platforms, shown in Figure~\ref{fig:daily-referrer-platform}. The overall trend shows that Twitter and Facebook drive more traffic than Telegram and Gab. Daily traffic volume fluctuates and can be affected by external events. For example, Jim Hoft, founder of~\textit{TGP}, was suspended by Twitter on Februrary 6, 2021.~\footnote{\url{https://www.forbes.com/sites/ajdellinger/2021/02/06/twitter-suspends-gateway-pundit-jim-hoft/?sh=761c0cff3653}} The suspension is likely related to the decline of traffic from Twitter on that day. Other than several peaks in late February, the volume of traffic from Twitter and Facebook has continued to decline. This finding suggests that suspending social media accounts that spread low quality URLs may indeed be an effective way to reduce the spread of misinformation.

\subsection*{Finding 3: Visitors to the site are more likely to be from areas that voted for Donald Trump during the 2020 presidential election}
Our web traffic data records IP and city-level geo-location label for every request. To better understand what types of audiences visit~\textit{TGP}, and the audiences' political preferences, we leverage the geo-location information to answer two key questions: given the fact that articles published on~\gp are pro-Trump, pro-Republican Party, and often related to the 2020 US election~\cite{fakenews-2020-election}, (1) is~\gp more popular in counties that voted Trump? and (2) is~\gp more popular in Republican states, Democratic states, or Swing states? 

We assume that each unique IP address is one unique visitor, and each visitor is a voter during the 2020 US Presidential Election. In reality, our assumption might not always be true. For example, multiple people in a household can share the same IP, or one person can visit the site from multiple IP addresses, or the person who visits the site is not eligible to vote. Even though those limitations exist, IP address is the most accurate proxy to real human traffic in our dataset. IP is also commonly used in security research to generate threat intelligence from traffic logs~\cite{microsoft-measurement, cisco-umbrella}. Additionally, we also filter out cities that have fewer than 1,000 unique visitors, because those cities are too small to allow a safe margin of error. For example, an IP address that belongs to a small city might get erroneously assigned to a neighboring city. After the filtering our data set has 596 US cities.\\

\noindent
\textbf{Question 1: Is~\gp more popular in counties that voted Trump?} 

To answer this question, we first collect county-level 2020 US election results, including the total number of voters, number of voters who voted for Trump, and number of voters who voted for Biden. Then in our web traffic data set, we group number of visits per city into number of visits per county. Finally for each county, we calculate two metrics:
\begin{align*} 
\text{\% voters who visited GP} (y) &= \frac{\text{unique number of IP}}{\text{total number of voters}} \\ 
\text{\% voters who voted for Trump} (x) &= \frac{\text{\# voters who voted Trump}}{\text{total number of voters}}
\end{align*}

\begin{figure}[!h]
  \centering
  \includegraphics[width=\linewidth]{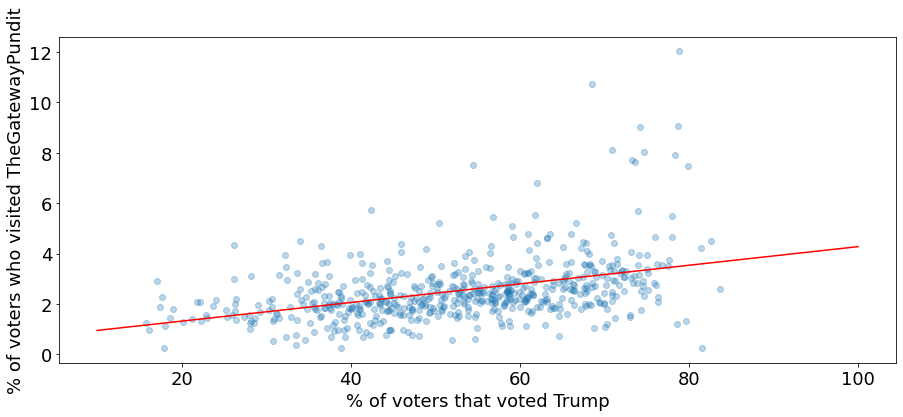}
  \caption{Scatter plot of city-level \% voters who voted for Trump (x-axis) in the 2020 US Election versus \% voters who visited~\gp (y-axis). The red line shows the expected value of $y$ given $x$, based on a linear regression model. The model has a slope of 0.037 and a r-squared value of 0.17, which means that $x$ and $y$ are positively correlated.}
  \label{fig:county-visit-vs-vote}
\end{figure}

Figure~\ref{fig:county-visit-vs-vote} shows the scatter plot of $x$ and $y$. The red line is the expected value of $y$ given $x$, based on a linear regression model. The r-squared value is 0.17 and the slope is 0.037, which indicates a positive correlation between \% voters who visited~\gp and \% voters who voted for Trump. Thus we do in fact find that~\gp is more popular in counties that voted Trump. This finding is consistent with a 2016 study that shows people from counties that voted for Trump are more likely to visit fake news sites~\cite{microsoft-measurement}.

\noindent\\
\textbf{Question 2: Is~\gp more popular in Republican states, Democratic states, or Swing states?}

On a state level, is there any correlation between a state's voting preference and the number of visits to~\textit{TGP}? To answer this question, we first assign each state to one of three categories, based on the voting result during the 2020 US Election:~\footnote{We use the classification from \url{https://swingleft.org/p/super-states}}

\begin{enumerate}[leftmargin=*]
    \item Republican. Most voters in this state tend to vote for Republican candidates. 18 states fall into this category.
    \item Democratic. Most voters in this state tend to vote for Democratic candidates. 20 states fall into this category.
    \item Swing States. 12 states fall into this category: Pennsylvania, Michigan, Arizona, Wisconsin, Georgia, Colorado, Texas, Florida, Ohio, North Carolina, Iowa, and Maine.
\end{enumerate}

We aggregate city-level counts into state-level counts. For each state, we calculate total and per-capita number of visitors. Per-capita number of visitors is defined as the total number of visitors divided by the population of the state. We then aggregate state-level counts based on voting preference, and calculate the mean and median number of visitors in Swing, Republican and Democratic states. Table~\ref{table:state-level-swing-rep-dem} shows the results.

Interestingly, we find that~\gp is more popular in Swing states and Republican states than in Democratic states. We perform pairwise t-test and find that the difference is statistically significant when using the per-capita count, but is not significant when using the absolute count. To understand why more people from Swing states visit the site, in the next section we analyze all articles published on the site during our data collection period. We show that most popular articles frequently mention topics related to Swing states, including ``2020 US Election fraud'', ``missing ballot'', or ``voting irregularity'', all of which are unverified or false claims. Those stories have more direct impact on people from Swing states than those from Republican or Democratic states.


\begin{table}
\begin{tabular}{p{0.39\linewidth}p{0.31\linewidth}p{0.22\linewidth}}
\toprule
voting preference & Mean number of visits (total,per-capita) & Median number of visits \\
\midrule
 Swing (12 states) &  4171,      0.025 &       2122,    0.020 \\
  Republican (18 states)&  3116,      0.022 &       1900,              0.019 \\
  Democratic (20 states) &  3096,      0.017 &       1652,              0.015 \\
\bottomrule
\end{tabular}
\caption{We group 50 states into three categories based on a state's voting preference. On average, more visitors come from Swing and Republican states than from Democratic states.}
\label{table:state-level-swing-rep-dem}
\end{table}

\subsubsection*{Visualizing hotspots.}
\begin{figure}[!h]
  \centering
  \includegraphics[width=0.9\linewidth]{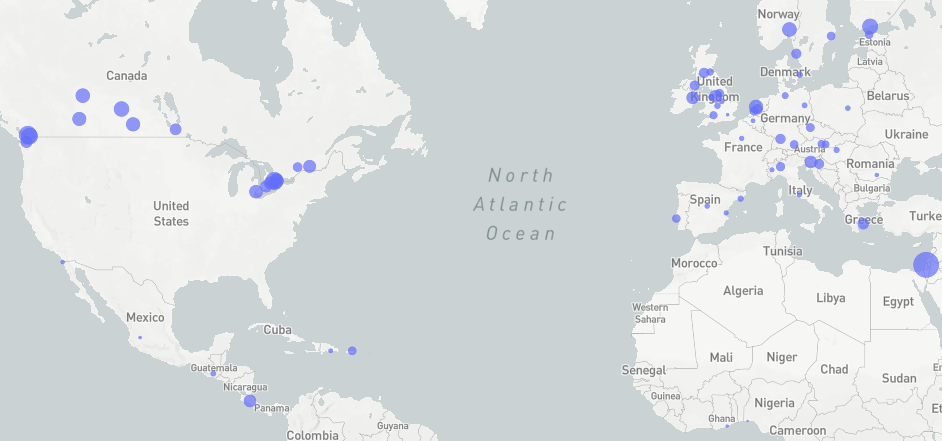}
  \caption{Top non-US cities based on \% visitors that visited \textit{TGP}. The larger a circle, the higher the percentage. Most non-US hotspots are located in Canada, Australia and Israel.}
  \label{fig:city-non-us}
\end{figure}

\begin{figure}[!h]
  \centering
  \includegraphics[width=0.9\linewidth]{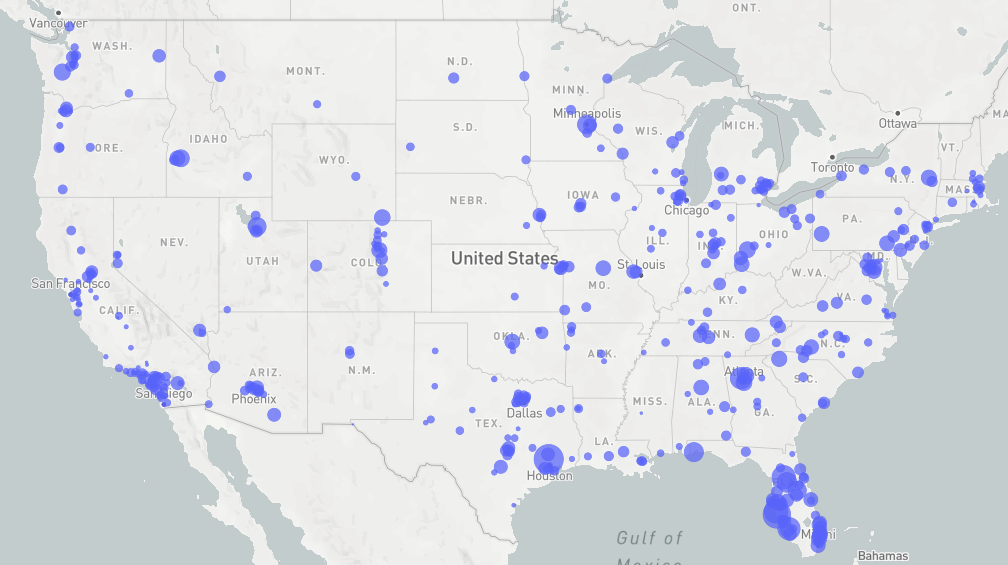}
  \caption{Top US cities based on \% visitors that visited~\gp. The larger a circle, the higher the percentage. Some cities belong to large metropolitan areas such as Denver, Houston and Chicago. There are also clusters of cities from Swing or Republican states such as Florida, Texas and Arizona.}
  \label{fig:city-us}
\end{figure}

To better understand where people visit \textit{TGP}, we visualize cities with a high concentration of visitors on two maps. We separate visits into two categories: those coming outside of the United States, and those coming from the United States. Figure~\ref{fig:city-non-us},~\ref{fig:city-us} show top US and non-US cities. The radius of each dot is in proportion to the percentage of city population that visited~\gp within a month. Most non-US visits come from cities in Canada, Australia, New Zealand and Israel. For visits within the United States, some come from metropolitan areas such as Denver, Houston and Chicago. Others come from cities within Swing or Republican states such as Florida, Texas and Arizona.

\subsection*{Finding 4: Topics related to ``election fraud'' receive more clicks and remain popular on the site for a longer period of time than other topics}
\label{subsection:topic-analysis}
During the one-month period of our study,~\gp published 1070 articles. Some stories go viral, others do not. What topics are discussed? What makes one topic goes viral? Is virality associated with the ``fakeness'' of the story? To better understand those connections, we use topic clustering technique to group~\gp articles into ten distinct topics. We then design two metrics to quantify the popularity of an article: number of unique visits (volume based), and number of minutes it takes to receive 50\%/90\%/95\% of all visits (time based). We first show the distribution of those two metrics over all articles, and then aggregate metrics into topics to identify viral content.\\

\noindent
\textbf{What topics are discussed?} 
Each article published on~\gp comes with a one-sentence title with references to key names and events. For example, one article published on February 18, 2021 is titled ``\textit{Maricopa County Audits Are Proving to Be a Waste of Time and Money, They Were Never Created to Identify the Suspected Election Fraud in the County}.'' Given the rich information from the title, we use non-negative matrix factorization (NMF) to cluster 1070 article titles into different topics. NMF is an unsupervised algorithm to extract topics from text corpus. In our case, the input to NMF is an article-word matrix, where each entry is the tf-idf weight of a word in an article. NMF then factorizes this matrix into a word-topic matrix, and a topic-article matrix. The number of topic is a user-defined parameter. After experimenting with different values, we set the parameter as 10, because the resulting topics are coherent and distinct among each other. Table~\ref{table:topic-cluster} shows keywords associated with each topic.\\

\begin{table}
\begin{tabular}{p{0.1\linewidth}p{0.9\linewidth}}
\toprule
topic &  keywords  \\
\midrule
 1 &  \textit{president donald trump acquittal } \\
 2 & \textit{joe biden kamala harris}   \\
 3  & \textit{2020 election fraud voter integrity}  \\
 4  &\textit{marjorie taylor greene, liz cheney}  \\
 5  &\textit{capitol riot antifa police fbi}  \\
 6  &\textit{governor andrew cuomo new york}  \\
 7  & \textit{democrat impeachment trial}   \\
 8  & \textit{maricopa arizona county ballot shredded dumpster}  \\
 9  &\textit{covid 19 vaccine virus cdc}   \\
 10  &\textit{dominion voting machine}  \\
\bottomrule
\end{tabular}
\caption{Keywords associated with each topic. We use non-negative matrix factorization to cluster 1070 articles into 10 topics.}
\label{table:topic-cluster}
\end{table}

\begin{figure}[!h]
  \centering
  \includegraphics[width=0.7\linewidth]{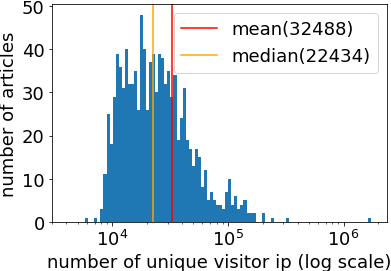}
  \caption{Histogram of unique visits per article. We use IP address as a proxy for visitor. The mean number of visits is 32,488, and the median is 22,434. This discrepancy suggests that some articles receive very high number of visits.}
  \label{fig:hist-article-visit}
\end{figure}

\noindent
\textbf{What topics receives more visits?} We first use the number of unique visit to measure article virality. Each unique IP address counts as one unique visit. Figure~\ref{fig:hist-article-visit} shows the histogram of number of unique visits per article. Overall, the average number of visits per article is 32,488; the median number of visits is 22,434. The distribution is skewed to the right, suggesting that some articles have very high number of unique visits.

\begin{figure}[!h]
  \centering
  \includegraphics[width=\linewidth]{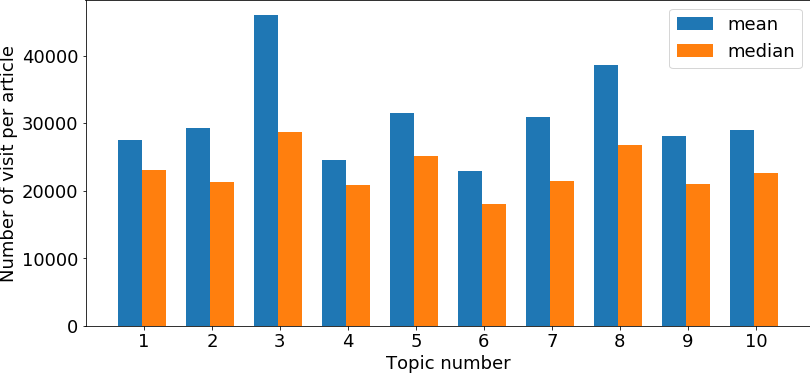}
  \caption{Mean and median number of unique visits per topic. The most visited topics are both related to the US 2020 election. Topic \#3 is related to election fraud, and topic \#8 is related to maricopa county ballot. Both claims are unverified conspiracy theories.}
  \label{fig:topic-distribution}
\end{figure}

We then aggregate article-level number of visits into topic-level. Figure~\ref{fig:topic-distribution} shows the mean and median number of visits per topic. The most visited topic is \#3, which according to Table~\ref{table:topic-cluster} is related to ``2020 US election fraud'', an unverified claim pushed by far-right news media. The second most visited topic is \#8, which covers ``voting irregularity and ballot counting in Maricopa county'', another unfounded claim. The popularity of those topics indicate that readers of \textit{TGP} had a huge appetite for articles about electoral fraud stories. The fact that those articles are published and remain popular three month after the US 2020 election shows that this type of misinformation can have a long-lasting effect on readers, and that misinformation does not have to cover real-time topics to remain popular.\\

\begin{figure}[!h]
  \centering
  \includegraphics[width=0.8\linewidth]{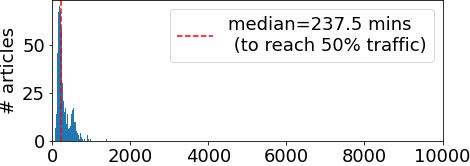}

  \includegraphics[width=0.8\linewidth]{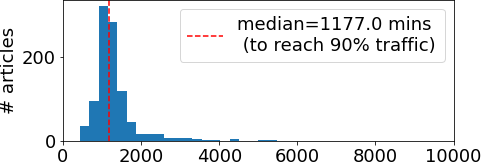}

  \includegraphics[width=0.8\linewidth]{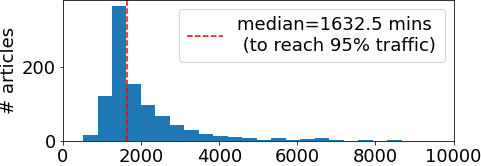}
  \caption{Histogram of number of minutes to reach 50\%, 90\% and 95\% of total unique visits. On average, an article receives 50\% of all visit traffic within 4 hours (280 minutes) of publication, and receives 90\% of all visit traffic within 20 hours (1,200 minutes) of publication.}
  \label{fig:hist-article-visit-length}
\end{figure}

\noindent
\textbf{Do viral topics last longer?} To quantify the popularity of an article in the time dimension, we measure how long it takes an article to receive 50\% ($t_1$), 90\% ($t_2$), and 95\% ($t_3$) of all visits, since the publication of the article. Figure~\ref{fig:hist-article-visit-length} shows that in median values, 50\% of visits come from the first 237 minutes (4 hours); 90\% of visits come from the first 1,177 minutes (20 hours), and 95\% of visits come from the first 1,634 minutes (28 hours). In general, it is rare to have an article that stays viral for more than a day.

 
\begin{figure}[!h]
  \centering
  \includegraphics[width=\linewidth]{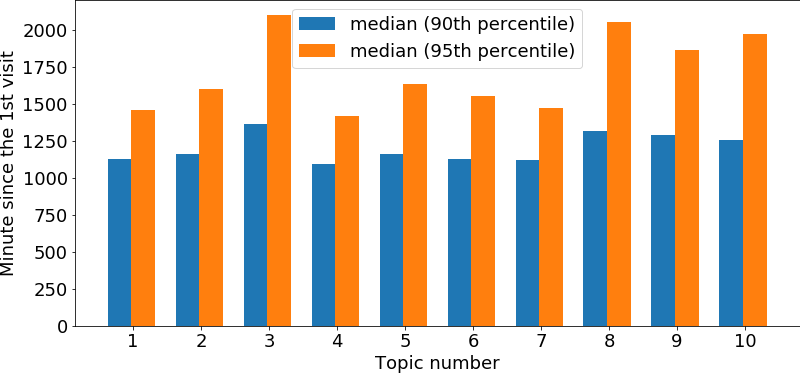}
  \caption{Median number of minutes to reach 90\% and 95\% of total visits per topic. Topics that trend longer include \#3 (US election fraud), \#8 (Arizona county ballot), and \#10 (Dominion voting machine). Topics that trend relatively shorter include \#1 (president donald trump acquittal) and \#7 (democrat impeachment trail). In general, conspiratorial topics last longer, while topics that report a known fact do not last as long.}
  \label{fig:topic-distribution-length}
\end{figure}

We then aggregate article-level count to topic-level count. Figure~\ref{fig:topic-distribution-length} shows the median $t_2$ and median $t_3$ for each topic. Topics that trend longer include \#3 (``US election fraud''), \#8 (``Arizona county ballot''), and \#10 (``Dominion voting machine''). Topics that trend shorter include \#1 (``President Donald Trump acquittal'') and \#7 (``Democrat Impeachment Trial''). The longest-trending topics are also the most visited topics. This suggests that viral topics are not only read by more people, but also last for a longer period of time. In general, topics related to conspiracy theories are more popular, while topics that state a known fact are less viral.

\section{Comparing web traffic data with social media engagement signals}
\label{section:comparison}
As we mentioned previously, existing research on news consumption mostly focuses on how news URLs are shared on social media platforms, especially Twitter and Facebook. While social media signals can tell us how people \textit{share} news, they do not answer how many people actually \textit{visit} each URL. Is there any correlation between social media sharing behavior and actual news consumption behavior? If so, how strong is the correlation? To answer those questions, we first collect Facebook and Twitter metrics that measure popularity of~\gp links shared on each platforms. We then test the correlational strength of different social media metrics against website visit count, and identify metrics that are good estimations of actual web visit counts. 

\subsection{Collecting posts from Facebook and Twitter}
Among 1070 online articles published by~\gp during our one-month data collection, 1020 received more than 10,000 unique web visits. To ensure the stability of our experiment, we focus on those 1020 URLs and discard URLs with lower web visit counts. We use Crowdtangle API to collect Facebook posts that contain any one of the 1020 URLs published by~\textit{TGP}. Crowdtangle is a data intelligence service that tracks aggregated engagements and interactions of posts from Facebook pages and groups (both public and private)~\cite{crowdtangle-api}. We use Twitter Academic API~\cite{twitter-api} to collect all original and public tweets that contain any one of the 1020 URLs. For each URL, we calculate seven metrics, shown in Table~\ref{table:metric-source}.

\begin{table}[!h]
\begin{tabular}{p{0.6\linewidth}p{0.35\linewidth}}
 \hline
 metric & source \\
 \hline
 number of unique visits (all) & web traffic dataset \\
 \# unique visits (from facebook.com) & web traffic dataset \\
 \# unique visits (from twitter.com) & web traffic dataset \\
 total number of FB reactions & Crowdtangle API \\
 total number of FB interactions & Crowdtangle API \\
 total number of likes & Twitter API \\
 total number of retweets & Twitter API\\
\hline
\end{tabular}
\caption{We calculate seven metrics to quantify the popularity of an article URL. We later correlate web traffic-based metrics with social media-based metrics.}
\label{table:metric-source}
\end{table}


\subsection{Measuring correlations}


We calculate Pearson correlations between each social media metric and (a) the number of visits from all traffic and (b) the number of visits from platform-specific traffic. Pearson correlation is the normalized covariance between two variables, and is used to summarize the strength of the linear relationship between two variables~\cite{pearson-correlation}. 

\begin{figure}
\centering
\subfloat{
  \includegraphics[width=0.49\linewidth]{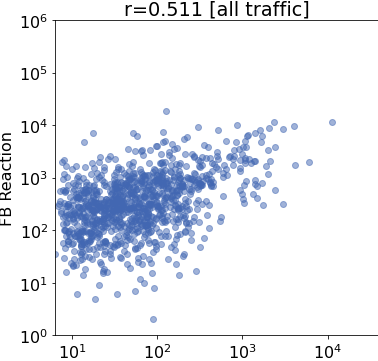}
}
\subfloat{
  \includegraphics[width=0.49\linewidth]{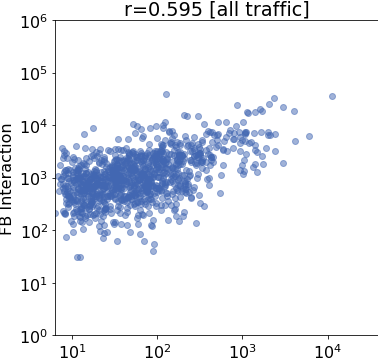}
}
\hspace{1mm}
\subfloat{
  \includegraphics[width=0.49\linewidth]{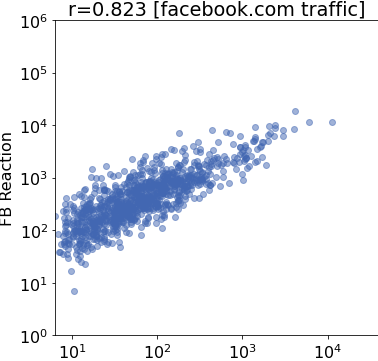}
}
\subfloat{
  \includegraphics[width=0.49\linewidth]{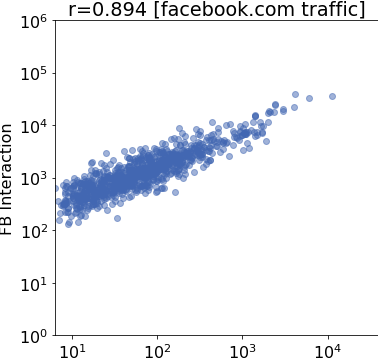}
}
\caption{Pearson correlations between Facebook engagement metrics (y-axis) and web visit count (x-axis). Both axes are in log scale. Top row is based on all traffic, and bottom row is based on traffic only from facebook.com. In each scatter plot, a dot represents a unique~\gp article URL.}
\label{fig:correlation-scatter-facebook}
\end{figure}

We first observe that Facebook metrics correlate better with traffic that only originated from Facebook than traffic originated from all sites. The same is true for Twitter metrics. For example, Figure~\ref{fig:correlation-scatter-facebook} shows that the Pearson correlation between total Facebook interaction and number of visits from~\url{facebook.com} is 0.894, while the correlation is only 0.595 for number of visits from all sites. Since social media metrics cannot capture URL sharing activities outside of the platform, the correlation significantly decreases when using number of visits from all sites.

\begin{figure}
\centering
\subfloat{   
  \includegraphics[width=0.49\linewidth]{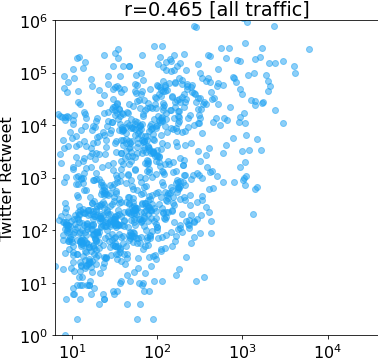}
}
\subfloat{
  \includegraphics[width=0.49\linewidth]{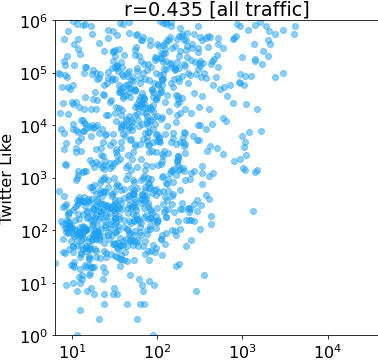}
}
\hspace{1mm}
\subfloat{   
  \includegraphics[width=0.49\linewidth]{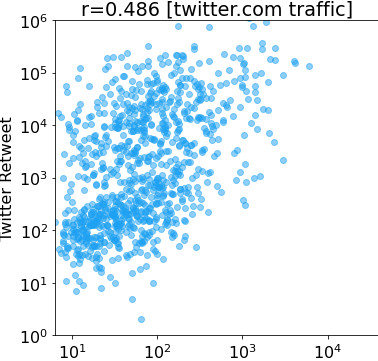}
}
\subfloat{
  \includegraphics[width=0.49\linewidth]{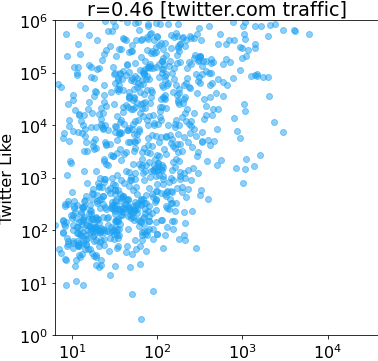}
}
\caption{Pearson correlations between Twitter engagement metrics (y-axis) and web visit count (x-axis). Both axes are in log scale. Compared with Facebook metric, Twitter metrics have weaker correlation with actual web visit count.}
\label{fig:correlation-scatter-twitter}
\end{figure}

We also observe that Facebook metrics correlate better with web visit count than Twitter metrics. For example, when we compare the first row of Figure~\ref{fig:correlation-scatter-facebook} against the first row of Figure~\ref{fig:correlation-scatter-twitter}, we see that Facebook interactions have a higher correlation with web visit counts than Twitter likes. The former metric has a Pearson correlation of 0.595 while the latter has a correlation of 0.435. Why is there a discrepancy? One reason could be that Facebook metrics we collect count both private and public posts, while Twitter metrics only count public posts. Understanding what other factors affect the correlation is a subject for future research.

To summarize, we validate that all social media metrics have a positive correlation with web visit counts. Therefore it is reasonable to use social media engagement signals as a proxy for URL popularity. However, there are limitations when using social media metrics, as they only capture link sharing activities on one platform. Given those insights, researchers should carefully choose from which platform to collect data and which engagement signals to use, as each metric has varying correlational strength. In the future, we plan to test more metrics to further understand correlations between how people share news on social media versus how people actually read news.


\section{Related Work}
\label{section:related-work}
Measuring who consumes and how people consume fake news is an important but challenging research area. Previous work mostly studies the spread of fake news on social media platforms~\cite{fakenews-2020-election}. For example, ~\cite{spread-of-true-and-false-news} collects tweets containing links to fake news sites, and concludes that fake news spread faster and further than traditional news. In another study using Twitter data,~\cite{fakenews-2016-election} claims that ``fake news accounted for nearly 6\% of all news consumption, but it was heavily concentrated'' on a small percentage of users. Similarly,~\cite{less-than-you-think}, \cite{cracking-open-the-news-feed} collect Facebook posts to understand news consumption behavior, and find that older people are more susceptible and share more fake news. 

While social media engagement signals can tell us how people share news on different platforms, they do not necessarily translate into web traffic to the news site~\cite{fake-news-not-antitrust}. One way to bridge this gap is to directly gather data from volunteers via browsing extensions. For example,~\cite{misinformation-in-action} asked participants to install a browser extension to measure their exposure to fake news. However, this approach is usually expensive and the sample size is small. 

To understand population-level news consumption behavior, there is an urgent need to collect ``unique datasets with increased validity''~\cite{tackling-misinformation}. Web traffic data is a direct measurement of news consumption. In one study, ~\cite{rise-and-fall-of-fake-news-sites} assesses user engagement by collecting traffic data from tracking services such as SimilarWeb and CheckPageRank.~\cite{microsoft-measurement} gathers browsing data from Microsoft Internet Explorer and Edge, and analyzes visitor patterns to a list of fake news domains before the 2016 US Election. 

Different from all previous approaches, we focus on collecting the entire web traffic to a single but important news site (\textit{TGP}). Our data set enables us to validate and extend previous traffic-based analysis. As far as we know, we are the first to test correlations between social media engagement signals and web traffic counts, by combining Twitter and Facebook posts with web traffic data.





\section{Discussion and Conclusion}
\label{section:conclusion}
In this paper, we collect and analyze a unique website traffic data set that contains more than 68 million visits to\textit{ The Gateway Pundit (TGP)}, a major far-right website known to spread fake news and conspiracy theories. We find that search engines and social media platforms are the main drivers that bring traffic to the site. Our geo-location analysis reveals that~\gp is more popular in counties that vote for Donald Trump, and our topic analysis shows that conspiratorial stories are more viral. Finally, we compare engagement signals derived from Twitter and Facebook posts with actual website visit counts, and find varying degrees of correlations. Our population-level behavioral analysis can help researchers design robust intervention methods to counter the spread of misinformation.

One major difficulty encountered during our research was our inability to analyze other comparable web traffic data sets. We reached out to several organizations who could offer such a data set, but did not move forward due to insufficient response. In the future, we plan to collaborate more with industry partners that have direct access to population-level news consumption data. Potential collaborators include web tracking companies and Internet service providers. As misinformation spreads over multiple platforms with an increasing speed, researchers need to be able to access more direct measurement data to quantify and understand the phenomenon of how people access low quality news websites.


\bibliography{bibliography}


\end{document}